\documentstyle[12pt,epsfig]{article}

\newcommand{\be}{\begin{equation}}
\newcommand{\ee}{\end{equation}}
\begin{document}
\topmargin 0pt
\oddsidemargin=-0.4truecm
\evensidemargin=-0.4truecm
\renewcommand{\thefootnote}{\fnsymbol{footnote}}
\newpage
\setcounter{page}{1}
\vspace*{-2.0cm}
\begin{flushright}
hep-ph/0007029
\end{flushright}
\begin{center}
{\Large \bf Reply to ``Is $U_{e3}$ really related to the solar neutrino 
solutions?''}
\vspace{0.8cm}

{\large E. Kh. Akhmedov\footnote{On leave from National Research Centre 
Kurchatov Institute, Moscow 123182, Russia. 
E-mail: akhmedov@cfif.ist.utl.pt}, 
G. C. Branco\footnote{E-mail: d2003@beta.ist.utl.pt} 
and M. N. Rebelo\footnote{E-mail: rebelo@beta.ist.utl.pt} }\\
\vspace{0.05cm}
{\em Centro de F\'\i sica das Interac\c c\~oes Fundamentais (CFIF)} \\
{\em Departamento de F\'\i sica, Instituto Superior T\'ecnico }\\
{\em Av. Rovisco Pais, P-1049-001, Lisboa, Portugal }\\
\end{center}
\vglue 0.6truecm
\begin{abstract}
In a recent paper \cite{ABR} we showed that, assuming no fine tuning
between certain elements of the neutrino mass matrix, one can 
link the element $U_{e3}$ of the lepton mixing matrix to solar and 
atmospheric neutrino oscillation parameters. 
This result has been recently criticized by Haba and Suzuki \cite{HS}. 
In the present note we show that their criticism is not valid and 
just reflects their failure to understand the content of our paper. 
\end{abstract}
\renewcommand{\thefootnote}{\arabic{footnote}}
\setcounter{footnote}{0}

In a recent paper \cite{ABR} we showed that, assuming no fine tuning
between certain elements of the neutrino mass matrix, one can use the
solar and atmospheric neutrino data to predict (up to a factor of the
order of unity) the leptonc mixing parameter $U_{e3}$. 
Conversely, a measurement of $U_{e3}$ in atmospheric or long baseline
accelerator or reactor neutrino experiments would help discriminate  
between possible oscillation solutions of the solar neutrino problem.
This result has been recently criticized by Haba and Suzuki (HS) \cite{HS}. 
In the present note we show that their criticism is not valid.

HS disputed our result on the grounds that no prediction regarding the 
value of $U_{e3}$ can be made basing on the solar and atmospheric neutrino 
oscillation parameters alone, without an additional nontrivial assumption. 
Obviously, in the absence of a theory of flavour, neutrino masses and
lepton mixing angles are all independent parameters and therefore knowledge 
of some of them does not allow one to predict the others unless additional 
assumptions are made. The necessity of an additional nontrivial assumption
has been clearly stated in our paper \cite{ABR}, and the assumption that we 
used was explicitly presented there: namely, no fine tuning between the 
elements $m_{12}$  and $m_{13}$ of the neutrino mass matrix $m_L$ in
the basis where the mass matrix of charged leptons has been diagonalised. 
This assumption allowed us to obtain relations between the values of
$U_{e3}$ and solar and atmospheric neutrino oscillation parameters, and 
therefore its nontriviality cannot be doubted. What HS actually dispute is
whether our assumption is justified\footnote{The authors of \cite{HS} claim 
to dispute the nontriviality of our assumption, but this appears to be a
language problem.}. Our results were based on the simple observation that, 
for any two arbitrary real numbers, the ratio of their sum and difference
is of the order of unity unless the absolute values of these numbers are 
finely tuned to be equal or nearly equal to each other. Our assumption barred 
the latter possibility for the elements $m_{12}$ and $m_{13}$ of $m_L$. 
Any fine tuning between the elements of the neutrino mass matrix in a 
physically meaningful basis is only natural if it is enforced by a symmetry. 
The existence of flavour symmetries leading to $m_{12}\simeq \pm m_{13}$ is 
certainly a possibility; we therefore stressed in \cite{ABR} that our 
predictions give just the likely values of $U_{e3}$. 

HS argue that our assumption leads to a relation between certain parameters 
$\alpha$ and $\beta$ introduced by them, which is not guaranteed by the data 
and therefore is not justified. 
Indeed, no relation between $\alpha$ and $\beta$ follows from the
experiment; such a relation was obtained in our paper through the use 
of the above stated assumption. HS have apparently missed this point. 

We have pointed out in \cite{ABR} that future experiments can test our 
predictions and so the assumption on which they are based. 
If the predicted relationships are not confirmed, this would signify the 
existence of a flavour symmetry leading to a fine tuning between 
$m_{12}$ and $m_{13}$, which by itself would be an interesting result. 

To summarize, the criticism of our paper \cite{ABR} by HS is based on
their misunderstanding of our analysis and is not valid.

\end{document}